\documentclass{pasj00}

\begin{document}
\SetRunningHead{Aoki \& Honda}{Enrichment of Pb in the Galactic Halo}
\Received{2007/0X/XX}
\Accepted{200X/XX/XX}

\title{Enrichment of lead (Pb) in the Galactic halo\thanks{Based on
    data collected at Subaru Telescope, which is operated by the
    National Astronomical Observatory of Japan. }}

 \author{%
  Wako \textsc{Aoki},\altaffilmark{1,2}
  Satoshi \textsc{Honda}\altaffilmark{3}
}

 \altaffiltext{1}{National Astronomical Observatory, Mitaka, Tokyo 181-8588}
 \email{aoki.wako@nao.ac.jp}

 \altaffiltext{2}{Department of Astronomical Science, The Graduate
  University for Advanced Studies, \\ Mitaka, Tokyo 181-8588}

 \altaffiltext{3}{Gunma Astronomical Observatory,
Takayama, Agatsuma, Gunma 377-0702}
\email{honda@astron.pref.gunma.jp}



\maketitle

\begin{abstract}

  We have determined lead (Pb) abundances for twelve red giants having
  [Fe/H] between $-2.1$ and $-1.3$, and its upper-limits for two lower
  metallicity objects, as well as the abundances of lanthanum (La) and
  europium (Eu). The averages of [Pb/Fe] and [Pb/Eu] are $-0.3$
  and $-0.6$, respectively, and no clear increase of these ratios with
  increasing metallicity is found. The [La/Eu] values are only
  slightly higher than that of the r-process component in
  solar-system material. Combining the previous studies for globular
  clusters, these results suggest little contribution of the s-process
  to Pb of the field stars studied here, supporting the estimate of Pb
  production by the r-process from the solar-system abundances.

\end{abstract}

\section{Introduction}\label{sec:intro}

Elements heavier than the iron group are mostly synthesized by
neutron-capture processes, which are distinguished into the rapid (r-)
and slow (s-) processes by the reaction time scale (e.g. Burbidge et
al.  1957).  The s-process fraction in solar-system material is
estimated by the classical or stellar s-process models calibrated for
the nuclei that are yielded only by the s-process (e.g. K{\"a}ppeler
et al. 1989, Arlandini et al. 1999). The r-process component is
derived by subtracting the s-process one from the solar-system
isotopic abundances.

The contributions of the s- and r-processes to Galactic objects
are estimated from the abundance ratios of elements that represent the
two processes (e.g. Ba/Eu, La/Eu; Truran et al. 1981; McWilliam et al.
1998; Burris et al. 2000). \citet{simmerer04} investigated the La/Eu
ratio for a large sample of metal-poor stars, and concluded that the
contribution of the s-process, with respect to the r-process,
increases with increasing metallicity, but large scatter of [La/Eu]
exists among stars having similar metallicity\footnote{[A/B] =
  $\log(N_{\rm A}/N_{\rm B})- \log(N_{\rm A}/N_{\rm B})_{\odot}$, and
  $\log\epsilon_{\rm A} = \log(N_{\rm A}/N_{\rm H})+12$ for elements A
  and B.}.

Another good probe of the s-process at low metallicity is Pb, one of
the heaviest elements including the isotope with magic numbers for
proton and neutron ($^{208}$Pb).  The Pb abundances have been measured
for many carbon-enhanced objects, which are believed to be affected by
mass transfer across binary systems from Asymptotic Giant Branch
(AGB) stars, and have been providing a useful constraint on the
s-process in low metallicity AGB stars.  In contrast,
measurements of Pb abundances for normal (non-carbon-enhanced) stars
are quite limited. \citet{travaglio01} argued enrichment of Pb in the
Galactic halo, thick and thin disks, reporting the Pb abundances for
several stars. Their sample, however, includes a star possibly
classified as a CH star, and the results for some objects are quite
uncertain.  \citet{ivans06} studied the heavy elements, including Pb,
for HD~221170 ([Fe/H]$=-2.2$) in detail. Pb abundances of stars in
globular clusters were recently determined by \citet{yong06} and
\citet{yong07}. 

This letter reports our abundance measurements of Pb and other
elements for bright metal-poor red giants, which would belong
to the Galactic halo. The results are compared with the solar-system
s- and r-process abundance ratios as well as that of
r-process-enhanced, extremely metal-poor stars.

\section{Observation, Analysis and Results}\label{sec:obs}

The sample of this study is selected from the list of
\citet{burris00}, who determined abundances of neutron-capture
elements like La and Eu, and that of \citet{beers00}, who studied
metallicity and kinematics of metal-poor stars. Bright stars
($V<10$) were selected for our purpose.


\begin{table}
\begin{center}
\caption{Object list}\label{tab:obj}
\begin{tabular}{lcrrr}
\hline
Object         & $(V-K)_{0}$ &  $U$ & $V$ & $W$ \\ 
\hline
BD+1$^{\circ}$2916      & 3.025 &  85 & $-$232 & 14 \\ 
BD+6$^{\circ}$648       & 2.950 &  $-$165 & $-$296 & 76 \\ 
BD+30$^{\circ}$2611     & 2.979 &  $-$10 & $-$59 & $-$284 \\ 
HD~3008         & 3.201 &  64 & $-$141 & 54 \\ 
HD~29574(V* HP Eri) & 2.927  & $-$296 & $-$213 & $-$223 \\ 
HD~74462        & 2.562 &  $-$121 & $-$284 & 70 \\ 
HD~141531       & 2.912 &  $-$189 & $-$278 & $-$67 \\ 
HD~204543       & 2.716 &  $-$11 & $-$130 & 26 \\ 
HD~206739       & 2.495 &  82 & $-$107 & $-$64 \\ 
HD~214925       & 3.269 &  129 & $-$512 & 136 \\ 
HD~216143       & 2.908 &  $-$329 & $-$268 & 78 \\ 
HD~220838       & 2.857 &  $-$27 & $-$76 & 22 \\ 
HD~221170       & 2.433 &  $-$119 & $-$123 & $-$22 \\ 
HD~235766       & 2.396 &  $-$305 & $-$260 & $-$4 \\ 
\hline
\end{tabular}
\end{center}
\end{table}

High resolution spectra were obtained for 14 stars given in
table~\ref{tab:obj} with the High Dispersion Spectrograph (HDS;
Noguchi et al. 2002) of the Subaru Telescope during several observing
runs including a short program in 2007. The resolving power
($R=\lambda/\delta \lambda$) of our observations ranges between 50,000
and 90,000, and the wavelength coverage is 4000--6800~{\AA},
3700--6400~{\AA} or 3100--4800~{\AA}, depending on observing runs. The
data reduction was made with the standard procedure using the
IRAF\footnote{IRAF is distributed by the National Optical Astronomy
  Observatories, which is operated by the Association of Universities
  for Research in Astronomy, Inc. under cooperative agreement with the
  National Science Foundation.} echelle package (e.g. Aoki et al.
2005). The signal-to-noise ratios of the spectra at the wavelength of
the Pb {\small I} line used in the analysis (4058~{\AA}) are 85 (per
0.024~{\AA} pixel) or higher.

Chemical abundance analyses were performed using the grid of model
atmospheres of \citet{kurucz93} with no convective overshooting
\citep{castelli97}\footnote{http://wwwuser.oat.ts.astro.it/castelli/grids.html}.
The effective temperatures ($T_{\rm eff}$) were estimated from
photometric colors (SIMBAD and 2MASS; Skrutskie et al. 2006) using
the scale of \citet{alonso99}.  The reddening was estimated from the
dust map of \citet{schlegel98} and the interstellar Na D line
absorption \citep{munari97} when available. We gave a priority to the
$(V-K)_{0}$ given in table~\ref{tab:obj} in the effective temperature
determination, while other colors were also investigated for
comparison purposes. The surface gravity ($g$), micro-turbulence
($v_{\rm turb}$), and metallicity ([Fe/H]) are determined by the LTE
analysis for Fe {\small I} and Fe {\small II} lines, as to obtain no
correlation of Fe abundances with line strengths and ionization
stages. For the analysis of most objects, we adopted the line list of
\citet{ivans06}. We added Fe {\small I} and {\small II} lines from
\citet{obrian91}, \citet{bard91}, and \citet{fuhr88} for relatively
metal-rich stars, for which only spectra of the UV--blue range are
available.  Although a slightly lower $\log g$ value than zero is
derived for HD~3008 and HD~214925, we adopted $\log g = 0.0$ for these
two stars. This little affects the derived abundance ratios of
neutron-capture elements. The stellar parameters adopted
for the following analyses are given in table 2.  The details of the
determinations of stellar parameters will be reported in our future
paper (Aoki et al., in preparation). The carbon abundances are
estimated from the CH band at 4223~{\AA}. The [C/Fe] of our sample
ranges between $-0.2$ and $-0.9$, indicating these stars are not
carbon-enhanced objects (table~\ref{tab:abund}).

\begin{figure}
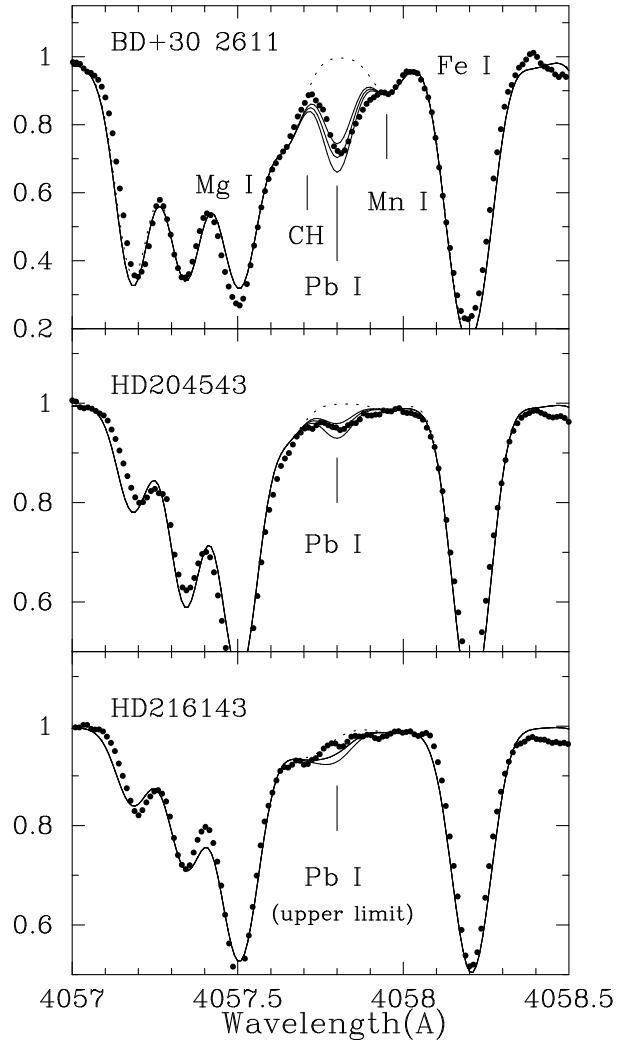

  \begin{center}
    \FigureFile(80mm,150mm){fig1.ps}
  \end{center} 

  \caption{ Comparisons of synthetic spectra for the region including
    the Pb I 4058~{\AA} line with the observed spectrum. The Pb
    abundances assumed in the calculations  are
    $\log \epsilon$ (Pb) $=+0.43 \pm 0.10$ (BD+30$^{\circ}$2611), $0.00 \pm
    0.15$ (HD~204543) and $-0.1, -0.3$ (HD~216143) for solid lines, 
    and $-\infty$ for dotted lines. The contamination of the CH line 
    at 4057.7~{\AA} is not severe in the spectra of our non-carbon-enhanced stars. 
}\label{fig:sp}
\end{figure}

The Pb abundances were determined by fitting synthetic spectra to the
observed Pb {\small I} 4058~{\AA} line. The line list of
\citet{simons89} including hyperfine splitting and isotope shifts,
which was used by \citet{aoki02}, was adopted, assuming the solar Pb
isotope ratios. We note that the derived abundance is
not sensitive to this assumption. Examples of the
observed and synthetic spectra are shown in figure 1.

The La and Eu abundances are determined using the line lists of
\citet{lawler01a} and \citet{lawler01b}, respectively (see also the
appendix of Ivans et al. 2006), including hyperfine splitting. The Eu
isotope ratio ($^{151}$Eu/$^{153}$Eu) is assumed to be unity. The La
abundance is determined from more than ten lines for all objects,
while the Eu lines used are dependent on the targets, because of the
differences of the wavelength coverage and the line strengths. The Eu
abundances are determined from lines in the red region (6049~{\AA},
6436~{\AA}, and/or 6645~{\AA}) for most stars. Two lines in the blue
region (4129~{\AA} and 4205~{\AA}) are used for objects for which only
blue spectra are available. We confirmed that the Eu abundances
derived from the blue and red lines agree within 0.13~dex for
HD~220838 and HD~221170. The results of the La, Eu and Pb abundances
($\log \epsilon$ values) are given in table~\ref{tab:res}, and the
abundance ratios are given in table~\ref{tab:abund}.

The random errors of the derived abundances are estimated as $\sigma
N^{-1/2}$, where $\sigma$ is the standard deviation of the derived
abundances from individual lines and $N$ is the number of lines
used. The $\sigma$ values for Fe and La lines are $\sim 0.1$~dex or
smaller. We adopted $\sigma$ of Fe {\small I} lines as that of Eu, for
which five lines or less are available for abundance analyses. The
random errors of the Pb abundances are estimated from the
fitting of the Pb {\small I} line, including the uncertainty of
continuum placement. For stars showing a clear Pb {\small I} feature
(the top panel of figure~\ref{fig:sp}), 0.10~dex is adopted as the
fitting error, while 0.15~dex is adopted for stars having weaker
features (the middle panel of figure \ref{fig:sp}). No clear feature
is found for HD~216143 (bottom panel of figure 1) and BD+6$^{\circ}$648, for
which we determined upper limits of Pb abundances.

Errors due to the uncertainty of atmospheric parameters are estimated
by calculating abundances changing atmospheric parameters by
$\Delta(T_{\rm eff})$=100~K, $\Delta(\log g)$=0.3~dex,
$\Delta$([Fe/H])=0.2~dex, and $\Delta(v_{\rm turb})=0.2$~km~s$^{-1}$
for BD+30$^{\circ}$2611.  These errors are added, in quadrature, to
the random errors estimated above, and are given in
table~\ref{tab:res}. The effects of the changes of atmospheric
parameters on the abundance ratios of Pb/La, Pb/Eu, and La/Eu are also
calculated, and are included in the errors given in table 3. We note
that the Pb abundance derived from the neutral species is lower if
higher gravity is adopted, while the sign is opposite for La and Eu
abundances from ionized species. This results in larger errors in
[Pb/La] and [Pb/Eu] than those in [La/Eu].


A comparison of our abundance results for HD~221170 with those of
\citet{ivans06} shows 0.1-0.2~dex differences in the abundances of Fe,
La, Eu and Pb, which are, at least qualitatively, explained by the
differences of adopted atmospheric parameters ($\Delta T_{\rm eff} =
$90~K and $\Delta \log g = $0.5~dex).  The abundances of HD~141531 are
determined by the present work for the same spectrum as used by
\citet{yong06}. The results agree within 0.05~dex between the two
measurements, adopting almost the same stellar parameters.

\section{Discussion and concluding remarks}\label{sec:disc}

\begin{figure}
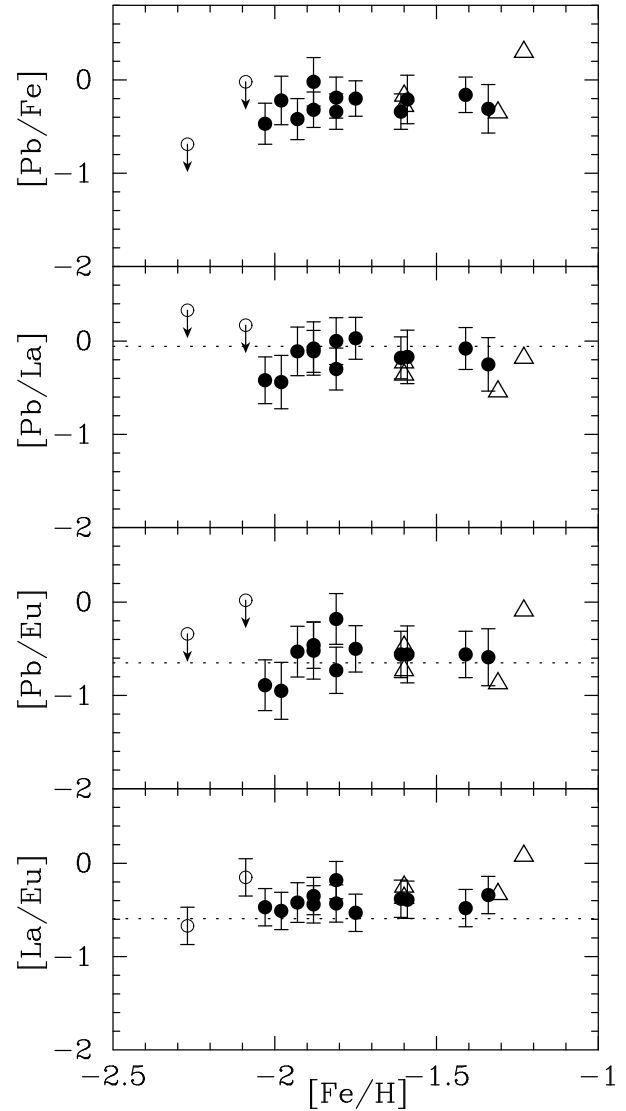

  \begin{center}
    \FigureFile(80mm,150mm){fig2.ps}
  \end{center} 
  \caption{Abundance ratios of the neutron capture elements La, Eu,
    and Pb as functions of [Fe/H]. Circles indicate our results. Open
    circles mean the result for the two objects for which only an
    upper limit of the Pb abundance is determined. Open triangles are
    the results for the globular clusters M4, M5, M13, and NGC~6752 by
    \citet{yong06} and \citet{yong07}. The dotted lines in the lower
    three panels show the abundance ratios of the solar-system
    r-process components \citep{simmerer04}.}\label{fig:pb}
\end{figure}

The top panel of figure \ref{fig:pb} shows [Pb/Fe] as a function of
[Fe/H] for the field stars of our sample and for globular cluster
stars. The Pb of our sample is underabundant, and the
abundance scatter is small: the average of [Pb/Fe] is $-0.27$ and the
standard deviation is 0.12~dex (table~\ref{tab:comp}). This result
indicates no rapid increase of Pb in this metallicity range 
in our sample. The open triangles in this panel present the averages of
Pb abundance ratios of stars in the globular clusters M~4, M~5, M~13
and NGC 6752 \citep{yong06, yong07}. While the [Pb/Fe] of three
clusters agrees with that found for our field stars, the value of
M~4 ([Fe/H]$=-1.23$) is higher by 0.5~dex.  The high Pb
abundance ratio should be the result of the large s-process
contribution to this cluster \citep{ivans99}.

In order to investigate the origins of Pb in the field stars, we 
depict the abundance ratios of Pb/La, Pb/Eu and La/Eu in the lower
three panels of figure \ref{fig:pb}. The dotted lines in these panels
mean the abundance ratios of the r-process component in solar-system
material (table~\ref{tab:comp}) determined by \citet{simmerer04}.  The
Pb/Eu and La/Eu of the s-process component are much higher than those
of the r-process component. Hence, the agreement of [Pb/Eu] of the
field stars with that of the solar-system r-process component implies
that the r-process is the dominant source of Pb in our sample, and the
contribution of the s-process is small if any.

The little contribution of the s-process to our sample is also
suggested by the La/Eu ratios. The La and Eu abundances of a large
sample of metal-poor stars were investigated by \citet{simmerer04}.
Seven stars among our objects are included in their sample, and the
[La/Eu] values measured by the two studies agree well. However, their
sample also includes stars having relatively high La/Eu ratios
([La/Fe]$\sim -0.2$) in this metallicity range, and they concluded the
existence of a spread in the level of s-process enrichment.
Measurements of Pb for the stars having higher La/Eu ratios are
desirable. It should be noted, however, that most of such stars in
their sample have relatively high temperature, and measurements of Pb
abundances are relatively difficult.

\citet{simmerer04} argued a correlation between the La/Eu ratio and
the kinematic properties of the objects, suggesting that ``high
velocity stars'' having high $U$ and $W$, or low $V$, show low La/Eu
ratios.  The kinematics data of our objects taken from \citet{beers00}
are given in table~\ref{tab:obj}. They show the characteristics of
``high velocity stars'', with a possible exception of HD~220838,
supporting the arguments of \citet{simmerer04}.

The above discussion is based on the r-process component in
solar-system material estimated by \citet{simmerer04}. However, the Pb
productions by the s- and r-processes are still very uncertain.
Indeed, the detailed abundance studies for r-process-enhanced,
extremely metal-poor stars reported significantly lower Pb abundance
ratios \citep{plez04, frebel07}. For instance, CS~31082--001 has
[Pb/Eu] of $-1.28$ \citep{plez04, hill02}, 0.63~dex lower than that of
the solar-system r-process component (table~\ref{tab:comp}). If we
adopt this as the abundance ratio of the r-process contribution
(assuming Eu is entirely the r-process origin), only 20\% of Pb of
our sample ($<$[Pb/Eu]$>=-0.59$) is from the r-process and the
remaining 80\% is the s-process origin.  In contrast, the average of
[La/Eu] of our sample ($<$[La/Eu]$>=-0.41$) is only slightly higher
than that of CS~31082--001 ([La/Eu]$=-0.52$). If
similar estimate to Pb is applied, 22\% of La originates from the
s-process. These estimates results in the Pb/La ratios of the
``s-process component'' of the field stars to be [Pb/La]$_{\rm
  s}=<$[Pb/La]$>+\log(0.80/0.22)=0.38$.

Thus, an alternative interpretation for our result is that the
r-process yield of Pb is as low as that found in the
r-process-enhanced extremely metal-poor stars, while the Pb production
with respect to La by the s-process at low metallicity is quite
efficient. An efficient synthesis of $^{208}$Pb at low metallicity is
predicted by the s-process calculations for AGB models with $^{13}$C
pocket (e.g. Busso et al. 1999). However, the decrease of Pb/La ratios
(possibly by one dex) with increasing metallicity, which is expected
from such models, is not found in our sample.  Another difficulty of
this interpretation is the abundance ratios of M~4 stars. While the
high La and Pb abundances of this cluster indicate large contributions
of the s-process, the $<$[Pb/La]$>$ of this cluster
(table~\ref{tab:comp}) is significantly lower than the [Pb/La]$_{\rm
  s}$ estimated above. That is, assuming large contributions of the
s-process to our field stars requests an unnatural assumption that the
s-process yields are significantly different between field stars and
globular cluster objects.


Our conclusion here is that Pb, as well as other neutron-capture
elements, in a majority of halo stars in the metallicity range of [Fe/H]$<-1.3$
primarily originated from the r-process. This supports the Pb
production by the r-process estimated from solar-system abundances
(e.g. Burris et al.  2000; Simmerer 2004), and suggests that the Pb
abundance of r-process-enhanced, extremely metal-poor stars (e.g.
CS~31082--001) is anomalously low. Such diversity of Pb production by
the r-process is recently argued by \citet{wanajo07}. However, given
the fact that some stars having high La/Eu ratios are known in this
metallicity range, further measurements of Pb abundances for a larger
sample are strongly desired.


\begin{table}
\footnotesize
\begin{center}
\caption{Elemental Abundances Results}\label{tab:res}
\begin{tabular}{lccccccccccccc}
\hline
Object &  \multicolumn{4}{c}{model parameters} & & \multicolumn{8}{c}{abundances results} \\ 
 \cline{2-5} \cline{7-14}
       & $T_{\rm eff}$ & $\log g$ & $v_{\rm turb}$ & [Fe/H] & & $\log\epsilon$(Fe) & error & $\log\epsilon$(La) & error & $\log\epsilon$(Eu) & error & $\log\epsilon$(Pb) & error \\
\hline
BD+1$^{\circ}$2916 & 4200 & 0.4 & 1.8 & $-$1.9 && 5.57 & 0.18& $-$0.87 & 0.12 &$-$1.22 & 0.14 &$-$0.20 & 0.19\\
BD+6$^{\circ}$648  & 4400 & 0.9 & 2.2 & $-$2.1 && 5.36 & 0.19 &$-$0.95 & 0.12 &$-$1.50 & 0.16 &$<$0.0 & ...\\
BD+30$^{\circ}$2611& 4250 & 0.5 & 1.9 & $-$1.5 && 6.04 & 0.18 &$-$0.27 & 0.12 &$-$0.49 & 0.14 & 0.43 & 0.19\\
HD~3008    & 4150 & 0.0 & 2.4 & $-$1.8 && 5.52 & 0.19& $-$1.02 & 0.14 &$-$1.30 & 0.16 &$-$0.35 & 0.22\\
HD~29574   & 4250 & 0.3 & 1.9 & $-$1.8 && 5.64 & 0.18 &$-$0.63 & 0.12 &$-$0.90 & 0.14 &$-$0.15 & 0.19\\
HD~74462   & 4600 & 1.5 & 1.3 & $-$1.4& & 6.11 & 0.18 &$-$0.18 & 0.12 &$-$0.54 & 0.14 & 0.35 & 0.26\\
HD~141531  & 4300 & 0.7 & 1.6 & $-$1.7 && 5.84 & 0.18 &$-$0.55 & 0.12 &$-$0.87 & 0.14 & 0.05 & 0.19\\
HD~204543  & 4600 & 1.0 & 2.2 & $-$1.9 && 5.64 & 0.19 &$-$0.78 & 0.12 &$-$1.30 & 0.16 & 0.00 & 0.22\\
HD~206739  & 4600 & 1.5 & 1.6 & $-$1.6 && 5.86 & 0.18 &$-$0.41 & 0.12 &$-$0.72 & 0.18 & 0.20 & 0.26\\
HD~214925  & 4050 & 0.0 & 2.1 & $-$2.0 && 5.42 & 0.18 &$-$0.86 & 0.12 &$-$1.09 & 0.20 &$-$0.50 & 0.22\\
HD~216143  & 4450 & 0.8 & 2.3 & $-$2.3 && 5.18 & 0.19 &$-$1.21 & 0.12 &$-$1.24 & 0.15 &$<$-$0.10$ & ...\\
HD~220838  & 4300 & 0.6 & 1.8 & $-$1.8 && 5.70 & 0.18 &$-$0.76 & 0.12 &$-$0.93&  0.16 & 0.05 & 0.19\\
HD~221170  & 4600 & 1.5 & 1.9 & $-$2.1 && 5.47 & 0.18 &$-$0.54 & 0.12 &$-$0.73 & 0.14 &$-$0.20 & 0.26\\ 
HD~235766  & 4650 & 1.2 & 1.9 & $-$1.8 && 5.57 & 0.18 &$-$0.60 & 0.12 &$-$0.86 & 0.14 & 0.10&  0.26\\
\hline
\end{tabular}
\end{center}
\normalsize
\end{table}

\clearpage

\begin{table}
\footnotesize
\begin{center}
\caption{Elemental Abundances Ratios}\label{tab:abund}
\begin{tabular}{lcccccccccc}
\hline
Object    &  [Fe/H]    & [C/Fe]  &[Pb/Fe] & error & [Pb/La] & error & [Pb/Eu] & error & [La/Eu] & error \\
\hline
BD+1$^{\circ}$2916 & $-1.88$ & $-0.66$ &  $-0.32$ & 0.19 & $ -0.11$ & 0.22 & $ -0.46$ &   0.25 & $ -0.35$ &   0.20 \\
BD+6$^{\circ}$648  & $-2.09$ & $-0.15$ & $<-0.02$ & .... & $ <0.17$ & .... & $ <0.02$ &   .... & $ -0.15$ &   0.20 \\
BD+30$^{\circ}$2611& $-1.41$ & $-0.78$ &  $-0.16$ & 0.19 & $ -0.08$ & 0.22 & $ -0.56$ &   0.25 & $ -0.48$ &   0.20 \\
HD~3008    &   $-1.93$ & $-0.51$ &  $-0.42$ & 0.22 & $ -0.11$ & 0.26 & $ -0.53$ &   0.27 & $ -0.42$ &   0.21 \\
HD~29574   &   $-1.81$ & $-0.83$ &  $-0.34$ & 0.19 & $ -0.30$ & 0.22 & $ -0.73$ &   0.25 & $ -0.43$ &   0.20 \\
HD~74462   &   $-1.34$ & $-0.40$ &  $-0.31$ & 0.26 & $ -0.25$ & 0.29 & $ -0.59$ &   0.31 & $ -0.34$ &   0.20 \\
HD~141531  &   $-1.61$ & $-0.58$ &  $-0.34$ & 0.19 & $ -0.18$ & 0.22 & $ -0.56$ &   0.25 & $ -0.38$ &   0.20 \\
HD~204543  &   $-1.81$ & $-0.63$ &  $-0.19$ & 0.22 & $  0.00$ & 0.25 & $ -0.18$ &   0.27 & $ -0.18$ &   0.20 \\
HD~206739  &   $-1.59$ & $-0.25$ &  $-0.21$ & 0.26 & $ -0.17$ & 0.29 & $ -0.56$ &   0.31 & $ -0.39$ &   0.20 \\
HD~214925  &   $-2.03$ & $-0.81$ &  $-0.47$ & 0.22 & $ -0.42$ & 0.25 & $ -0.89$ &   0.27 & $ -0.47$ &   0.20 \\
HD~216143  &   $-2.27$ & $-0.37$ & $<-0.69$ & .... & $ <0.33$ & .... & $<-0.34$ &   .... & $ -0.67$ &   0.20 \\
HD~220838  &   $-1.75$ & $-0.49$ &  $-0.20$ & 0.19 & $  0.03$ & 0.22 & $ -0.50$ &   0.25 & $ -0.53$ &   0.20 \\
HD~221170  &   $-1.98$ & $-0.61$ &  $-0.22$ & 0.26 & $ -0.44$ & 0.29 & $ -0.95$ &   0.31 & $ -0.51$ &   0.20 \\
HD~235766  &   $-1.88$ & $-0.66$ &  $-0.02$ & 0.26 & $ -0.08$ & 0.29 & $ -0.52$ &   0.31 & $ -0.44$ &   0.20 \\
\hline
\end{tabular}
\end{center}
\normalsize
\end{table}


\begin{table}
\begin{center}
\caption{Abundance ratios of neutron-capture elements}\label{tab:comp}
\begin{tabular}{lccccc}
\hline
Object & Ref.\footnotemark[$*$]  & [Pb/Fe]  & [Pb/La]  & [Pb/Eu] & [La/Eu]  \\
\hline
field stars (12 objects)\footnotemark[$\dagger$]  & 1  & $-$0.27 (0.12) & $-$0.18 (0.15) & $-$0.59 (0.20) & $-$0.41 (0.09) \\
M~4 stars (12 objects)\footnotemark[$\dagger$] & 2 &  0.30 (0.07) & $-$0.18 (0.06) & $-$0.09 (0.09) &  0.08 (0.06) \\
s-process (solar-system) & 3 &    & 0.022 &  1.471 &  1.449    \\
r-process (solar-system) & 3 &    &$-$0.057 &  $-$0.650& $-$0.593    \\
CS~31082--001         & 4 & 0.35 & $-$0.76 & $-$1.28 & $-$0.52   \\
\hline
\multicolumn{6}{@{}l@{}}{\hbox to 0pt{\footnotesize
    \footnotemark[$*$] References: (1)This work; (2)\citet{yong07}; (3)\citet{simmerer04}; (4)\citet{hill02} and \citep{plez04}. }\hss} \\
\multicolumn{6}{@{}l@{}}{\hbox to 0pt{\footnotesize
    \footnotemark[$\dagger$] Averages and standard deviations of abundance ratios.   }\hss}
\end{tabular}
\end{center}
\end{table}


\begin{thebibliography}{}


\bibitem[Alonso, Arribas, Mart\'{i}nez-Roger
(1999)]{alonso99}Alonso, A., Arribas, S., \& Mart\'{i}nez-Roger,
C. 1999, \aaps, 140, 261

\bibitem[Aoki et al.(2002)]{aoki02} Aoki, W., Ryan, S.~G., 
Norris, J.~E., Beers, T.~C., Ando, H., \& Tsangarides, S.\ 2002, \apj, 580, 
1149 

\bibitem[Aoki et al. (2005)]{aoki05} Aoki, W., et al.\ 2005, 
\apj, 632, 611 

\bibitem[Arlandini et al. (1999)]{arlandini99} Arlandini, C.,
K\"{a}ppeler, F., Wisshak, K., Gallino, R., Lugaro, M., Busso, M., \&
Straniero, O. 1999, \apj, 525, 886

\bibitem[Bard, Kock, and Kock (1991)]{bard91} Bard, A., Kock, A., \&
Kock, M. 1991, \aap, 248, 315

\bibitem[Beers et al.(2000)]{beers00} Beers, T.~C., Chiba, M., 
Yoshii, Y., Platais, I., Hanson, R.~B., Fuchs, B., \& Rossi, S.\ 2000, \aj, 
119, 2866 

\bibitem[Burbidge et al.(1957)]{burbidge57} Burbidge, E.~M., 
Burbidge, G.~R., Fowler, W.~A., \& Hoyle, F.\ 1957, Reviews of Modern 
Physics, 29, 547 

\bibitem[Burris et al.(2000)]{burris00} Burris, D.~L., 
Pilachowski, C.~A., Armandroff, T.~E., Sneden, C., Cowan, J.~J., \& Roe, 
H.\ 2000, \apj, 544, 302 

\bibitem[Busso et al.(1999)]{busso99} Busso, M., Gallino, R., 
\& Wasserburg, G.~J.\ 1999, \araa, 37, 239 

\bibitem[Castelli et al.(1997)]{castelli97} Castelli, F., Gratton, 
R.~G., \& Kurucz, R.~L.\ 1997, \aap, 318, 841 
 

\bibitem[Frebel et al.(2007)]{frebel07} Frebel, A., Christlieb, 
N., Norris, J.~E., Thom, C., Beers, T.~C., \& Rhee, J.\ 2007, \apjl, 660, 
L117 

\bibitem[Fuhr, Martin, Wiese (1988)]{fuhr88} Fuhr, J.R., Martin,
G.A., \& Wiese, W.L. 1988, J. Phys. Chem. Ref. Data, 17, supple. 4

\bibitem[Hill et al.(2002)]{hill02} Hill, V., et al.\ 2002, 
\aap, 387, 560 

\bibitem[Ivans et al.(2006)]{ivans06} Ivans, I.~I., Simmerer, 
J., Sneden, C., Lawler, J.~E., Cowan, J.~J., Gallino, R., \& Bisterzo, S.\ 
2006, \apj, 645, 613 

\bibitem[Ivans et al.(1999)]{ivans99} Ivans, I.~I., Sneden, C., 
Kraft, R.~P., Suntzeff, N.~B., Smith, V.~V., Langer, G.~E., \& Fulbright, 
J.~P.\ 1999, \aj, 118, 1273 

\bibitem[K{\"a}ppeler et al.(1989)]{kappeler89} K{\"a}ppeler, F., 
Beer, H., \& Wisshak, K.\ 1989, Reports of Progress in Physics, 52, 945

\bibitem[Kurucz (1993)]{kurucz93} Kurucz, R.\ 1993, ATLAS9 
Stellar Atmosphere Programs and 2 km/s grid.~Kurucz CD-ROM No.~13.~ 
Cambridge, Mass.: Smithsonian Astrophysical Observatory, 1993., 13, 

\bibitem[Lawler, Bonvallet, Sneden (2001a)]{lawler01a} Lawler, J.~E., 
Bonvallet, G., \& Sneden, C.\ 2001a, \apj, 556, 452   

\bibitem[Lawler et al.(2001b)]{lawler01b} Lawler, J.~E., 
Wickliffe, M.~E., den Hartog, E.~A., \& Sneden, C.\ 2001b, \apj, 563, 1075 

\bibitem[McWilliam(1998)]{mcwilliam98} McWilliam, A.\ 1998, \aj, 
115, 1640

\bibitem[{Munari \& Zwitter(1997)}]{munari97}
Munari, U., \& Zwitter, T. 1997, A\&A, 318, 269

\bibitem[Noguchi et al. (2002)]{noguchi02} Noguchi, K., et al.\ 
2002, \pasj, 54, 855 

\bibitem[O'Brian et al.(1991)]{obrian91} O'Brian, T.~R., 
Wickliffe, M.~E., Lawler, J.~E., Whaling, J.~W., \& Brault, W.\ 1991, 
Journal of the Optical Society of America B Optical Physics, 8, 1185

\bibitem[Plez et al.(2004)]{plez04} Plez, B., et al.\ 2004, 
\aap, 428, L9 

\bibitem[Simons et al. (1989)]{simons89} Simons, J.W., Palmer, B.A.,
Hof, D.E., \& Oldenborg, R.C. 1989, J. Opt. Soc. Am. B., 6, 1097

\bibitem[Schlegel, Finkbeiner, Davis (1998)]{schlegel98} Schlegel,
D.J., Finkbeiner, D.P., \& Davis, M. 1998, \apj, 500, 525

\bibitem[Simmerer et al.(2004)]{simmerer04} Simmerer, J., Sneden, 
C., Cowan, J.~J., Collier, J., Woolf, V.~M., \& Lawler, J.~E.\ 2004, \apj, 
617, 1091 

\bibitem[Skrutskie et al.(2006)]{skrutskie06} Skrutskie, M.~F., et
al.\ 2006, \aj, 131, 1163

\bibitem[Travaglio et al.(2001)]{travaglio01} Travaglio, C., 
Gallino, R., Busso, M., \& Gratton, R.\ 2001, \apj, 549, 346 

\bibitem[Truran(1981)]{truran81} Truran, J.~W.\ 1981, \aap, 97, 
391 

\bibitem[Wanajo(2007)]{wanajo07} Wanajo, S.\ 2007, \apjl, 666, 
L77

\bibitem[Yong et al.(2006)]{yong06} Yong, D., Aoki, W., 
Lambert, D.~L., \& Paulson, D.~B.\ 2006, \apj, 639, 918 

\bibitem[Yong et al.(2007)]{yong07} Yong, D., Lambert, D.~L., 
Paulson, D.~B., \& Carney, B.~W.\ 2007, ApJ, in press,
arXiv:0710.2367 


\end{thebibliography}
\end{document}